\documentclass{ws-ijmpa}

\begin{document}

\markboth{A. K. Dash, D. P. Mahapatra and B. Mohanty}
{Expectation of forward-backward rapidity correlations in $p$+$p$ collisions at the  LHC energies}

\catchline{}{}{}{}{}
%
%%%%%%%%%%%%%%%%%%%%%%%%%%%%%%%%%%%%%%%%%%%%%%%%%%%%%%%%%%%%%%%%%%%%

\title{Expectation of forward-backward rapidity correlations in $p$+$p$ collisions at the  LHC energies
%INSTRUCTIONS FOR TYPESETTING 
%MANUSCRIPTS\footnote{For the title, try not to use more than 
%3 lines. Typeset the title in 10 pt roman, uppercase and 
%boldface.}
}

\author{Ajay Kumar Dash}

\address{Universidade Estadual de Campinas,\\
13083-970, Campinas, Sao PauloCity, Brazil\\
ajayd@ifi.unicamp.br}

\author{Durga Prasad Mahapatra}

\address{Institute of Physics,\\
Sachivalaya Marg, Bhubaneswar - 751005, India\\
dpm@iopb.res.in}

\author{Bedangadas Mohanty}

\address{Variable Energy Cyclotron Centre\\
1/AF, Bidhan Nagar, Kolkata - 700064, India\\
bmohanty@vecc.gov.in}

\maketitle

\begin{history}
\received{Day Month Year}
\revised{Day Month Year}
\end{history}

\begin{abstract}
Forward-backward correlation strength ($b$) as a function of
pesudorapidity intervals for experimental data from $p$+$\bar{p}$ non-singly diffractive
collisions are compared to PYTHIA and PHOJET model calculations. 
The correlations are discussed as a function of rapidity window
($\Delta \eta$) 
symmetric about the central rapidity as well as rapidity window
separated by a gap ($\eta_{gap}$)  between forward and backward regions.  While the
correlations are observed to be independent of $\Delta \eta$, it is
found to decrease with increase in $\eta_{gap}$. This reflects the
role of short range correlations and justifies the use of $\eta_{gap}$
to obtain the accurate information about the physics of interest, the long range correlations.
The
correlation strength from PYTHIA are in agreement with the available
experimental data while those from the PHOJET give higher values.
For $p$+$p$ collisions at $\sqrt{s}$ = 7, 10 and 14 TeV, the correlation
strength from PHOJET are lower compared to those from PYTHIA,
this is in contrast to the observations at lower energies.
%Thus providing useful                                                                                                                                                       
%information to distinguish between the two models at LHC energies.                                                                                                          
The experimental $b$ value shows a linear dependence on  $\ln \sqrt{s}$ with
the maximum value of unity being reached at $\sqrt{s}$ = 16 TeV, beyond the
top LHC energy. However calculations from the PYTHIA and PHOJET models
indicate a deviation from linear dependence on $\ln \sqrt{s}$ and
saturation in the $b$ values being reached
beyond $\sqrt{s}$ = 1.8 TeV. Such a saturation in correlation values could
have interesting physical interpretations related to clan structures in
particle production. Strong forward-backward correlations are associated
with cluster production in the collisions. The average number of charged particles
to which the clusters fragments, called the cluster size, are found to
also increase linearly with $\ln \sqrt{s}$ for both data and the models studied.
The rate of increase in cluster size vs. $\ln \sqrt{s}$ from models studied are
larger compared to those from the data
and higher for PHOJET compared to PYTHIA. Our study indicates that the
forward-backward measurements will provide a clear distinguishing observable
for the models studied at LHC energies.
\keywords{Forward-Backward correlations, Long range correlations,
  cluster formation, p+p collisions at LHC}
\end{abstract}

\ccode{PACS numbers: 1.200, 1.300}

\section{Introduction}	
Understanding the mechanism of particle production in $p$+$p$ collisions is
one of the first goals of the experiments at the Large Hadron Collider (LHC).
Several new results from the LHC experiments in terms of particle
multiplicities  at $\sqrt{s}$ = 0.9, 2.36, 2.76 and 7 TeV are being compared
to various models of particle production~\cite{ALICE,ALICE1,ALICE2,CMS,CMS1,ATLAS}.
The most popular models for comparison to data from $p$+$p$ collisions
being PYTHIA~\cite{pythia,pythia1,pythia2,pythia3} and PHOJET~\cite{phojet,phojet1}.
The PHOJET model combines the ideas based on a dual
parton model (DPM)~\cite{dpm,dpm1} on soft process of particle production and uses lowest-order
perturbative QCD for hard process. Regge phenomenology is used to parameterize the
total, elastic and inelastic cross-sections. The initial and final state parton shower
are generated in leading log-approximation. PYTHIA on the other hand uses string
fragmentation as a process of hadronization and tends to use the perturbative
parton-parton scattering for low to high $p_{T}$ particle production. Initial results
have shown that both models do not have perfect agreement with multiplicity measurements
at LHC energies studied so far~\cite{ALICE,ALICE1,ALICE2,CMS,CMS1,ATLAS}. However it may be mentioned
that several of these models
in turn are used to obtain various correction factors for the experimental measurements.
In this paper we suggest that the correlations between the particles produced in forward
and backward rapidities can be used to discriminate between the various models of
particle production, more reliably at the LHC energies.

The forward-backward correlations previously observed had several physical interpretations.
The correlations over small range in rapidity are believed to be dominated by short-range
correlations as due to resonance decays, those occurring over large rapidity range could
be interpreted to be due to multiple parton interactions. In early 1985, a statistical
interpretation of the forward-backward correlations observed in ISR energies was
provided~\cite{carruthers}. The interpretation was based on clusters being produced
in these hadronic collisions according to a negative binomial distribution and the final
hadrons are a result of the decay of these clusters. An extended version of such a statistical
scenario can be found in Ref.~\cite{twostage}.  A more dynamical interpretation was
provided based on the DPM as in Ref.~\cite{dpm2}. The experimental data was also
interpreted in terms of a model based on a unitarized model which
included soft and semi-hard components (minijets). In such a model the average number of
particles produced was proportional to the effective number of inelastic collisions.
The increase of forward-backward correlations with $\sqrt{s}$
was fully generated by the superposition of the
different impact parameter contributions to the inelastic cross section~\cite{minijets}.
The extension of this interesting idea is to consider a weighted superposition of
two classes of events in hadronic collisions: soft and semi-hard processes. The behavior
of the semi-hard component on observed forward-backward correlations
can lead to interesting interpretation in terms of the clan
structure of particle production as discussed in Refs~\cite{clan}.
The authors of Refs~\cite{clan}
envision a possibility of formation of new species of clans at LHC and a possible
phase transition in clan production mechanism.

Recently the forward-backward correlations have been related to the
simplest form of partonic interaction that exhibits back-to-back correlation.
Such a formulation is found to give a fairly good description of
data from STAR Collaboration in $p$+$p$ collisions at $\sqrt{s}$ = 200 GeV~\cite{hwa}.
Studies of forward-backward correlations provide baseline measurements to look
for long range rapidity correlation in heavy-ion collisions. STAR experiment has
recently made this relative (comparison between $p$+$p$ and different collision centrality Au+Au collisions
at the center of mass energies of 200 GeV) measurements to claim the existence
of a large long range correlation in central Au+Au collisions~\cite{star}.
Although the experimental data cannot differentiate
whether the actual underlying mechanism is due to those in a
dual parton model~\cite{dpm,dpm1}
DPM or a Color Glass Condensate~\cite{cgc,cgc1}, both however require
that the long range correlations are produced by multiple parton-parton interactions.
Further it is argued that the clustering of
color sources could lead to forward-backward
correlations~\cite{colorsources}. Hence the measurement of the long
range forward backward
correlations in the multiplicity of produced particles in high energy
collisions will give us insight into the space-time
dynamics of the collision.

In this paper, we first review the existing experimental data on forward-backward correlations,
discuss the model calculations and methods used to extract these correlations.
This is followed by comparison of the experimental data on forward-backward correlations
with those from PYTHIA and PHOJET
models, expectations at top LHC energies and interpretation of the data in terms
of cluster production in high energy hadronic collisions. We find that the forward-backward
correlations in  $p$+$p$ collisions at $\sqrt{s}$ = 7, 10, 14 TeV will help differentiate
various models and hence the underlying particle production mechanism. Saturation of
correlation strength at LHC energies could indicate new physics.

\section{Experimental data, Model Simulation and Correlations}
The experimental charged particle data reviewed in this paper are from
the E735 and UA5 collaboration. These data are for $p$+$\bar{p}$ collisions and
corresponds to non-singly diffractive events. The E735 data~\cite{e735} corresponds
to $\sqrt{s}$ = 300, 546, 1000 and 1800 GeV while those from UA5~\cite{ua5} corresponds
to $\sqrt{s}$ = 200, 546 and 900 GeV. The pesudorapidity acceptance range
for the E735 experimental data is $\mid \eta \mid$ $<$ 3.25 and those
for UA5 is $\mid \eta \mid$ $<$ 4.0.

The model results presented in this paper are from PYTHIA (Version 6.4)
and PHOJET (Version 1.1, uses jetset74 from PYTHIA) with default settings. The event type selected
are non-singly diffractive as for the existing measurements. For comparison
to existing data the simulations are done for $p$+$\bar{p}$ collisions
with the experimental acceptances included. For predictions at LHC energies
of $\sqrt{s}$ = 10 and 14 TeV the model calculations are done for $p$+$p$
collisions. A realistic transverse momentum cut of above 100 MeV for
charged particles are applied.

The forward-backward correlations can be obtained by two methods.
\begin{itemize}
\item If $N_{b}$ is the multiplicity in backward hemisphere and
    $N_{f}$ is the corresponding multiplicity in forward hemisphere
    for the same event, then the correlation co-efficient can be obtained
    by plotting $\langle N_{b} \rangle$($N_{f}$) vs. $N_{f}$. Where $\langle .. \rangle$
    denotes event average. The resultant distribution can be fitted to a linear function as
    $\langle N_{b} \rangle$($N_{f}$) = $a$ + $b$ $N_{f}$, to obtained
    the correlation strength,$b$.

\item The correlation coefficient can also be defined as
      $b = \frac{\langle N_{f} N_{b} \rangle - \langle N_{f} \rangle \langle N_{b} \rangle}{\langle N_{f}^{2} \rangle - \langle N_{f}\rangle}^{2}$. Most of the results presented in this paper
uses this method for calculating $b$. For the model results presented we have
explicitly checked and found there is good agreement between the two methods.
\end{itemize}

\begin{figure}
\begin{center}
 \includegraphics[scale=0.6]{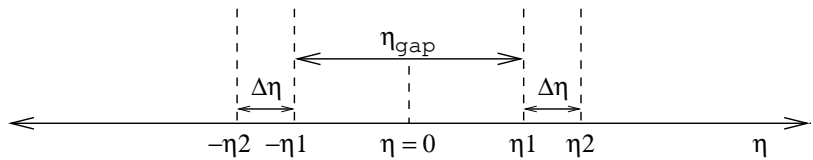}
\caption{Definition of $\Delta\eta$ and $\eta_{\mathrm {gap}}$ used in forward-backward correlation
studies.}
\label{defn}
\end{center}
\end{figure}

Before we proceed towards discussion of the experimental data and results from
model calculations, we discuss few relevant experimental aspects.
The forward-backward correlation results for a given $\sqrt{s}$ are usually presented
as a function $\Delta\eta$ and $\eta_{\mathrm {gap}}$. Fig.~\ref{defn} shows
how these quantities are defined. The $\Delta\eta$ corresponds to calculation
of correlation coefficient in a symmetric window about the central rapidity.
Increasing $\Delta\eta$ window includes the contributions from smaller
$\Delta\eta$ window. The results as a function of $\eta_{\mathrm {gap}}$ corresponds
to correlation co-efficient being calculated in some fixed $\Delta\eta$ window
separated by a gap in rapidity between forward and backward regions by an amount
$\eta_{\mathrm {gap}}$.

Most of the existing forward-backward correlation results at high energies
are from $p$+$\bar{p}$ collisions. The current work presents the predictions
of the correlation co-efficient from PYHTIA and PHOJET models in $p$+$p$
collisions at LHC energies and compares them to extrapolations of results
from $p$+$\bar{p}$ collisions. So it is essential to check using models if a difference is expected
in correlation values between $p$+$p$ and $p$+$\bar{p}$ collisions at a given
$\sqrt{s}$. Note however that the STAR Collaboration has measured the strength of 
charged particle forward-backward multiplicity correlation, $b$  in $p$+$p$ collision at 
$\sqrt{s}$ = 200 GeV~\cite{brijesh}. It is about 3 - 4 times smaller than the one measured 
by UA5 in $p$+$\bar{p}$ collision at the same energy apparently~\cite{ua5}. The eta gap in 
STAR is 0.2 where as in UA5 it is 0.5.  Also both the experimental acceptances are different. 
STAR has $\mid \eta \mid$ $<$ 1.0 where as UA5 has $\mid \eta \mid$ $<$ 4. 

\begin{figure}
\begin{center}
 \includegraphics[scale=0.4]{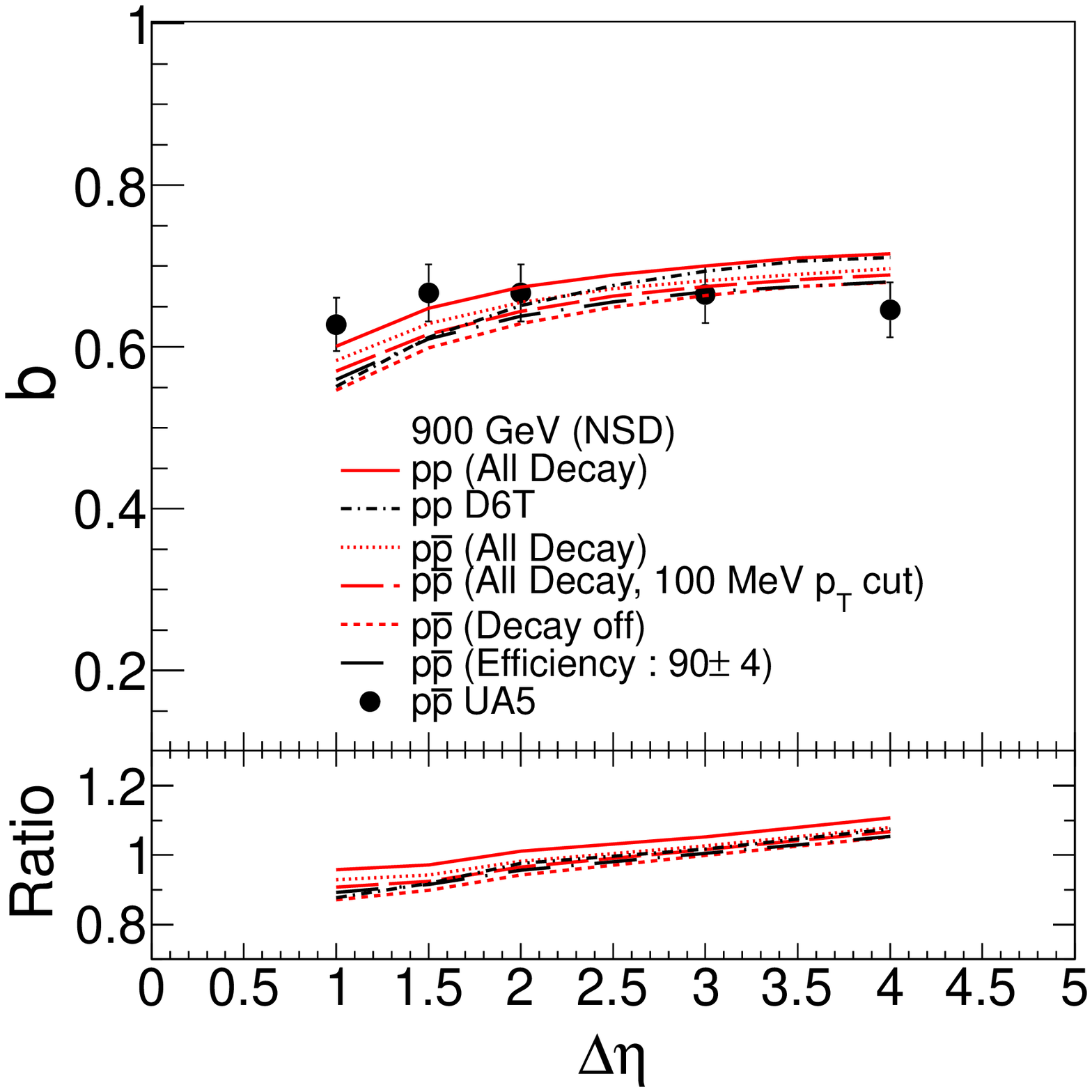}
 \includegraphics[scale=0.4]{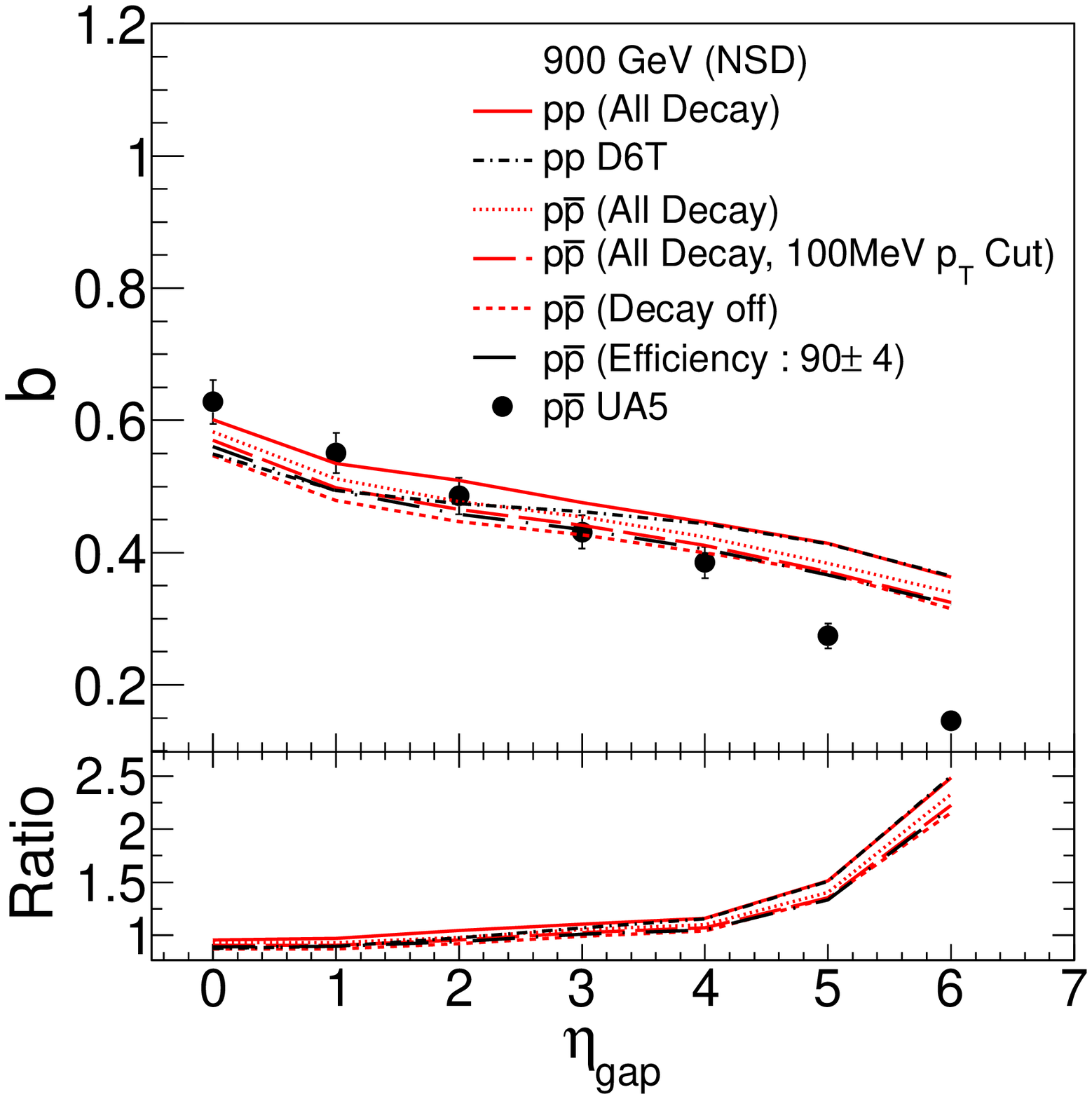}
\caption{Forward-backward correlation co-efficient from PYTHIA at $\sqrt{s}$ = 900 GeV as a function of $\Delta \eta$ and $\eta_{gap}$. 
The results are from non-singly diffractive events and presented for both $p$+$p$ and $p$+$\bar{p}$ collisions. The effect of 
weak decay, lower transverse momentum cut off of 100 MeV/$c$, typical charged particle detector efficiency, different model parameter 
are studied. For comparison the results from $p$+$\bar{p}$ collisions in UA5 experiment are also 
shown. The bottom panels of the figure shows the ratio of the
correlations from simulations with various conditions to those
measured by UA5 experiment.}
\label{study}
\end{center}
\end{figure}

Fig.~\ref{study} shows the correlation co-efficient ($b$) calculated
using non-singly diffractive events from PYTHIA model at $\sqrt{s}$ = 900 GeV
as a function of $\Delta\eta$ for $p$+$p$ and $p$+$\bar{p}$ collisions. No
appreciable difference is observed between the two colliding systems for
correlations studied as a function of $\Delta \eta$ and $\eta_{gap}$.
Also shown for comparison are the
charged particle correlation co-efficient results from UA5 experiment.
Usually due to experimental limitations the analysis is carried out
with a lower $p_{\mathrm T}$ cut off. As seen in Fig.~\ref{study} a
typical experimental cut-off of 100 MeV on $p_{\mathrm T}$ does not seem
to change the correlation values as a function of $\Delta \eta$ and $\eta_{gap}$.
We have also investigated the effect of weak decay
particle contributions to this analysis. The difference between weak decay
on and weak decay off for $p$+$\bar{p}$ collisions is noticeable for smaller
$\Delta\eta$  and the difference vanishes as we go to larger $\Delta\eta$.
This is along expected lines, as decay effect will introduce short-range
correlations. The effect of finite charged particle detection
efficiency has been studied by varying it between 86\% to 94\%, no
appericiable change is observed. The effect of a different tuned
version of PYTHIA (D6T) has also been investigated. This also does not
seem to affect the observed correlations by a large amount. The bottom
panel of the figure shows the ratio of the correlation from simulation with various
effects to the correlation measured by UA5 experiment. One observes
that for the study with respect to $\Delta\eta$ the variations lie
within 20\%. For the study with respect to the $\eta_{gap}$, the ratio
is close to unity upto a value of $\eta_{gap}$ ~ 4 units, then the
simulation starts to diverge away from the measurements.

\section{Correlations in $\Delta\eta$ and $\eta_{\mathrm {gap}}$}

\begin{figure}
\begin{center}
 \includegraphics[scale=0.35]{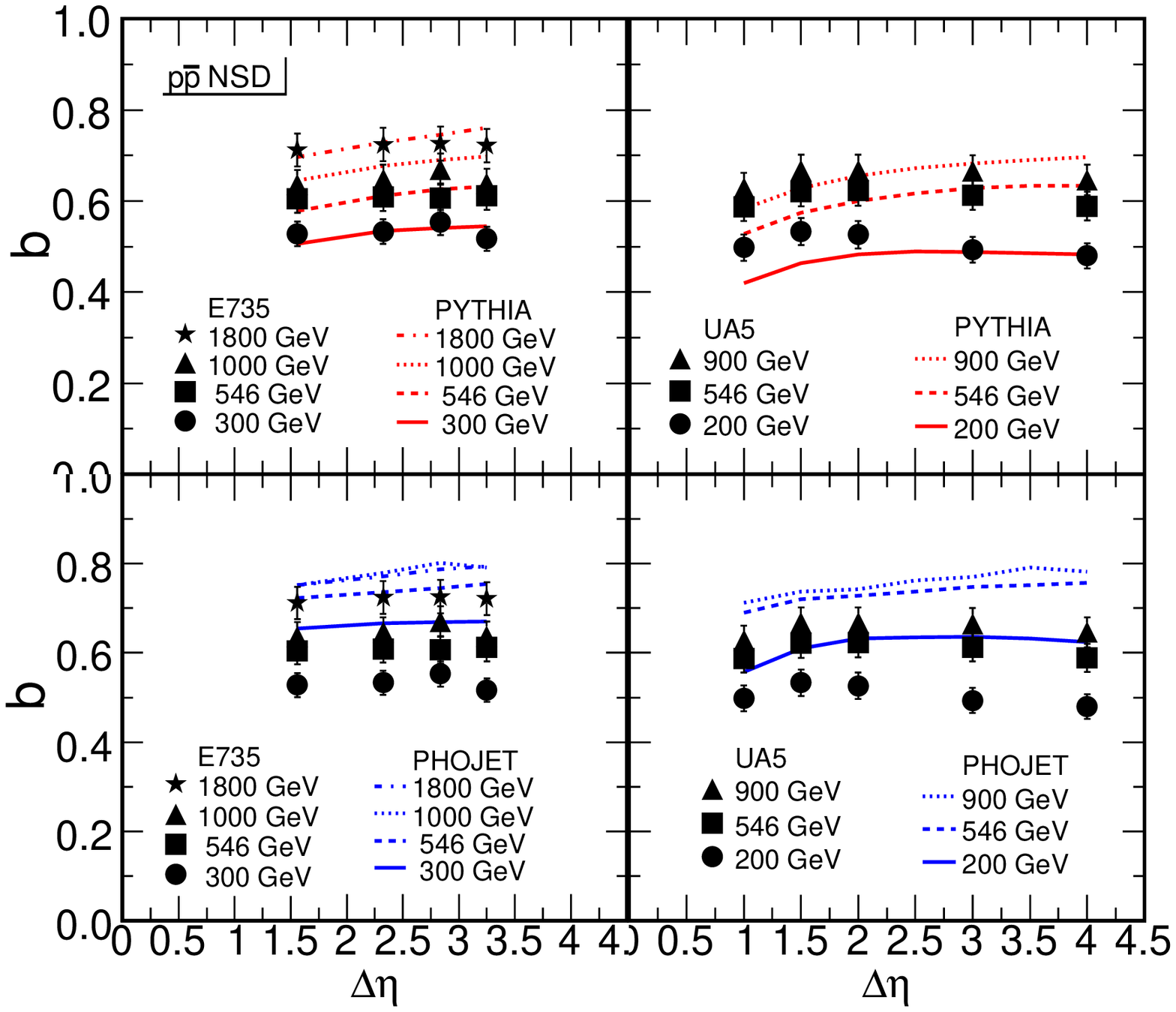}
\includegraphics[scale=0.35]{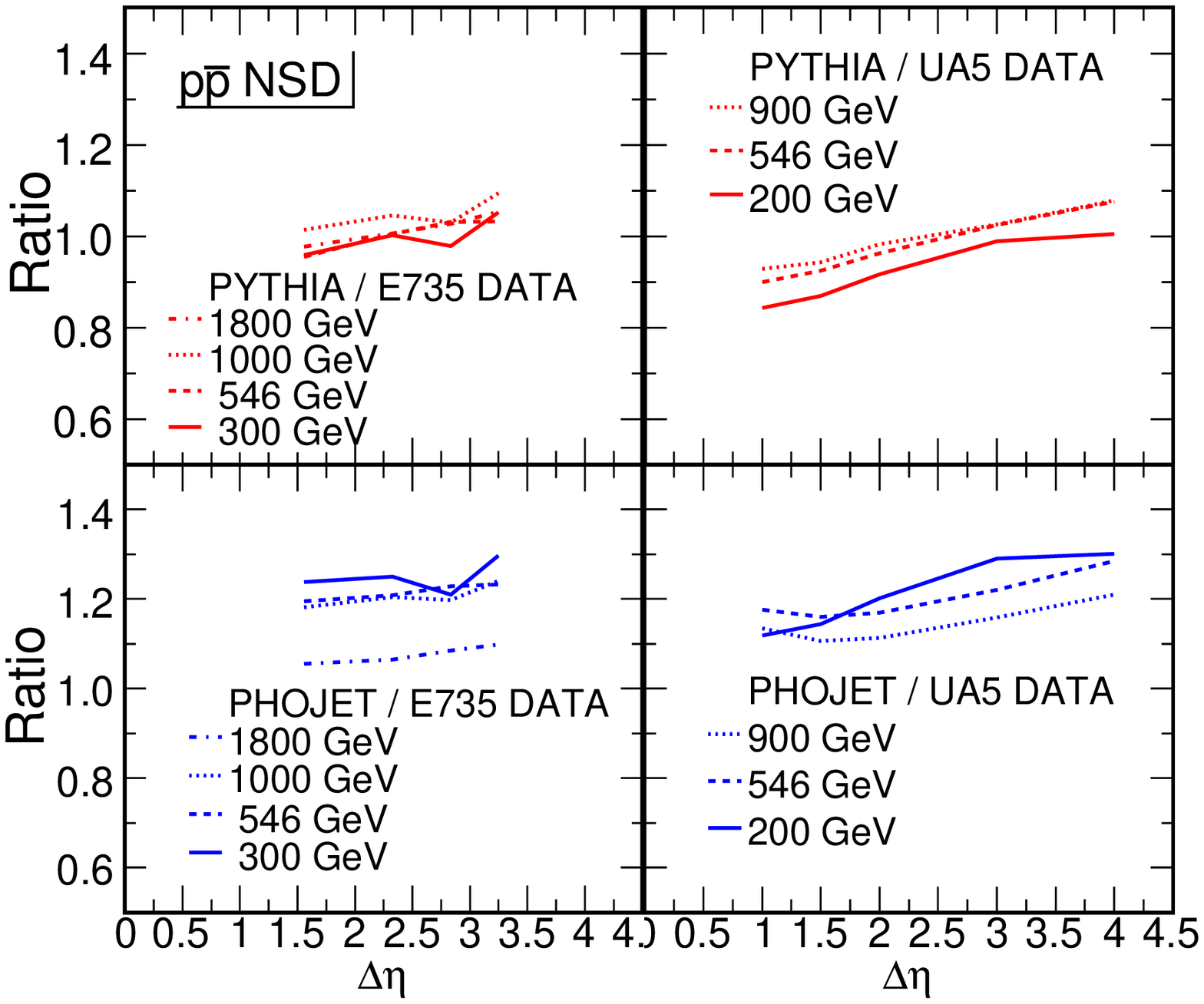}
\caption{Top panel: Forward-backward correlation strength ($b$) as a function of
$\Delta\eta$ for $p$+$\bar{p}$ non-singly diffractive
collisions at various $\sqrt{s}$ from E735 and UA5 experiments. The
measurements are compared to PYTHIA and PHOJET model calculations.
Bottom panel: The ratio of model calculations to experimental data.}
\label{delta_eta}
\end{center}
\end{figure}

Fig~\ref{delta_eta} shows the correlation strength ($b$) as a function of
$\Delta\eta$ for non-singly diffractive events in $p$+$\bar{p}$ collisions
from E735~\cite{e735} and UA5~\cite{ua5} experiments at various $\sqrt{s}$ compared to
corresponding results from PYTHIA and PHOJET model calculations. The following
observations can be made:
(a) The correlation strength both in experimental data and simulations are
    observed to be almost independent of $\Delta\eta$,
(b) correlations seem to increase with $\sqrt{s}$, and
(c) PYTHIA model calculations agrees well with the experimental data,
    while PHOJET tends to give higher correlations for all the measured
    $\sqrt{s}$. This can be seen from the ratio of the correlation
    value from model to data in the bottom panel of the Fig. ~\ref{delta_eta}.
As the larger $\Delta\eta$ includes contribution from
smaller $\Delta\eta$ intervals, such a data may not provide accurate
information about long range correlations. These correlations could
include significant short-range correlations.

\begin{figure}
\begin{center}
 \includegraphics[scale=0.35]{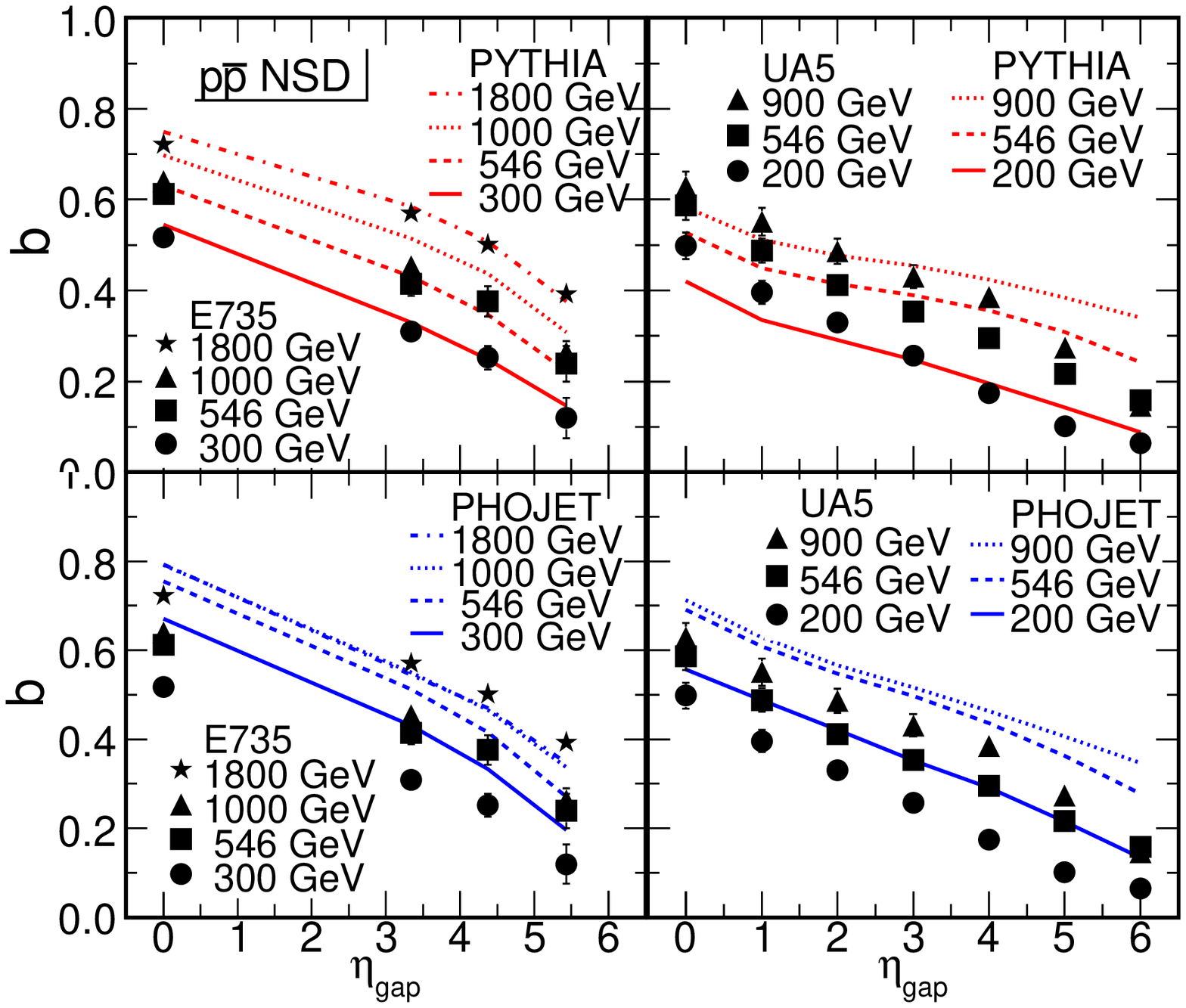}
\includegraphics[scale=0.35]{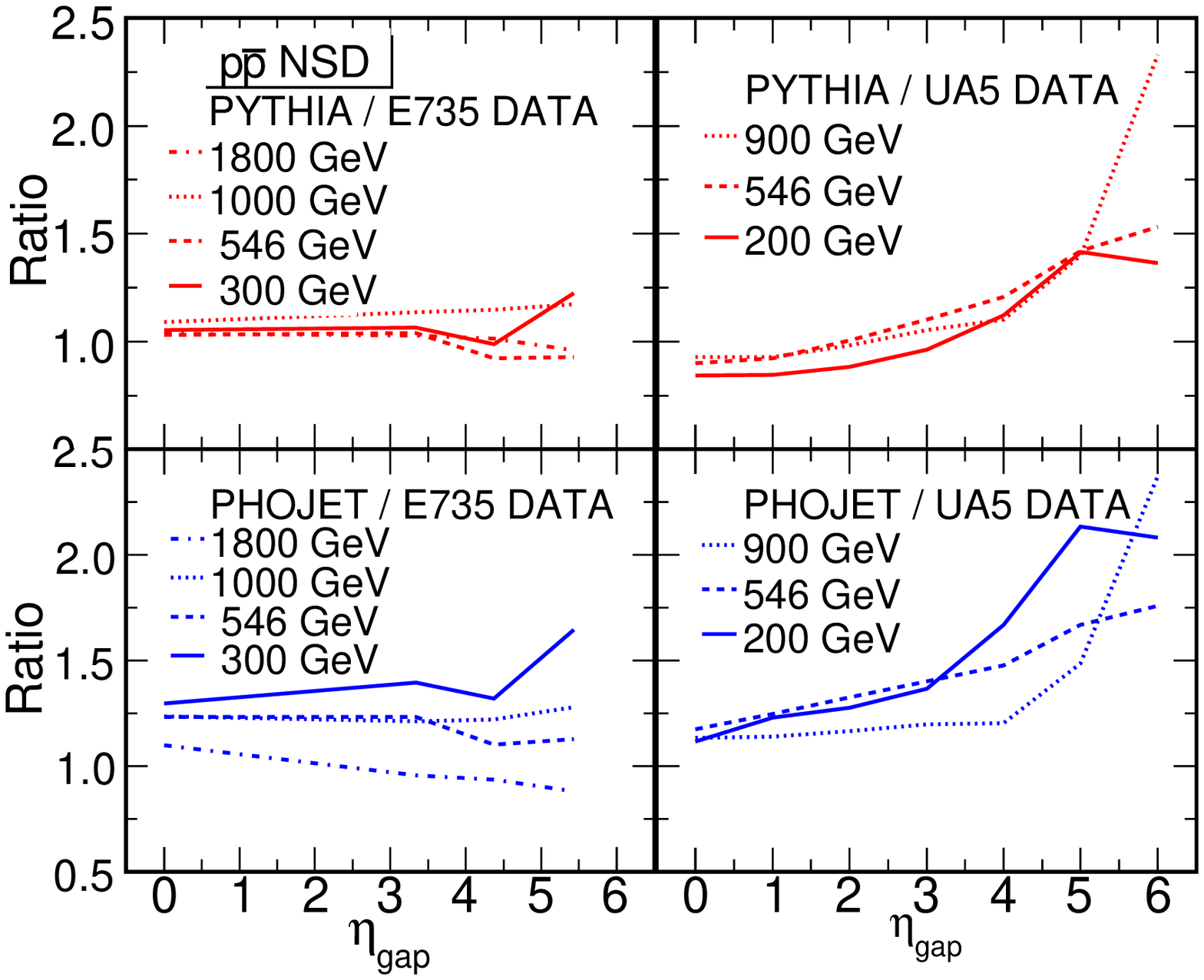}
\caption{Top panel: Forward-backward correlation strength ($b$) as a function of
$\eta_{\mathrm {gap}}$ for $p$+$\bar{p}$ non-singly diffractive
collisions at various $\sqrt{s}$ from E735 and UA5 experiments. The
measurements are compared to PYTHIA and PHOJET model
calculations. Bottom panel: The ratio of model calculations to experimental data. }
\label{eta_gap}       % Give a unique label                                                                                                                                  
\end{center}
\end{figure}

Fig~\ref{eta_gap} shows the correlation strength ($b$) as a function of
$\eta_{\mathrm {gap}}$. The value of $b$ decreases with increase in
$\eta_{\mathrm {gap}}$ indicating the diminishing contribution from
short-range correlations. With a  $\eta_{\mathrm {gap}}$ of around 2 units
still significant correlation at the level of 30 - 40\% are observed.
Both the models reproduce the decreasing trend of the correlations with
increasing $\eta_{\mathrm {gap}}$. PYTHIA model calculations has better
agreement with the measurements, while PHOJET model in general over estimates
the correlation strength at most of the $\sqrt{s}$, except for
$\sqrt{s}$ = 1.8 TeV.
This can be seen from the ratio of correlations from models to that
from data as shown in the bottom panel of Fig.~\ref{eta_gap}.
The agreement with the highest energy data by both the models poses the
question whether at LHC energies the current measurements will have the
distinguishing power. To investigate this possibility, we now discuss
the predictions at three LHC energies of $\sqrt{s}$ = 7, 10 and 14 TeV from
PYTHIA and PHOJET models.

\begin{figure}
\begin{center}
 \includegraphics[scale=0.35]{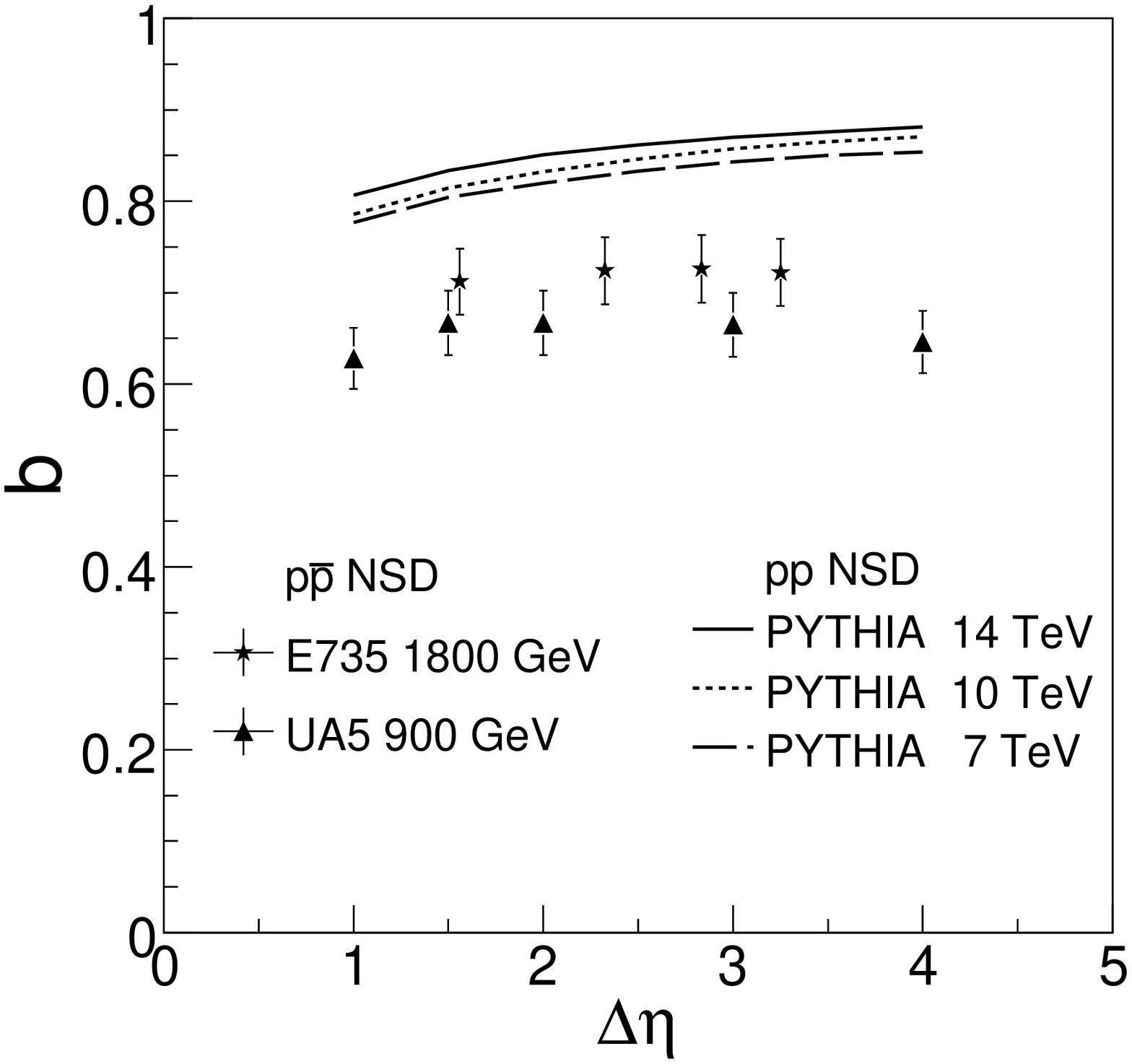}
 \includegraphics[scale=0.35]{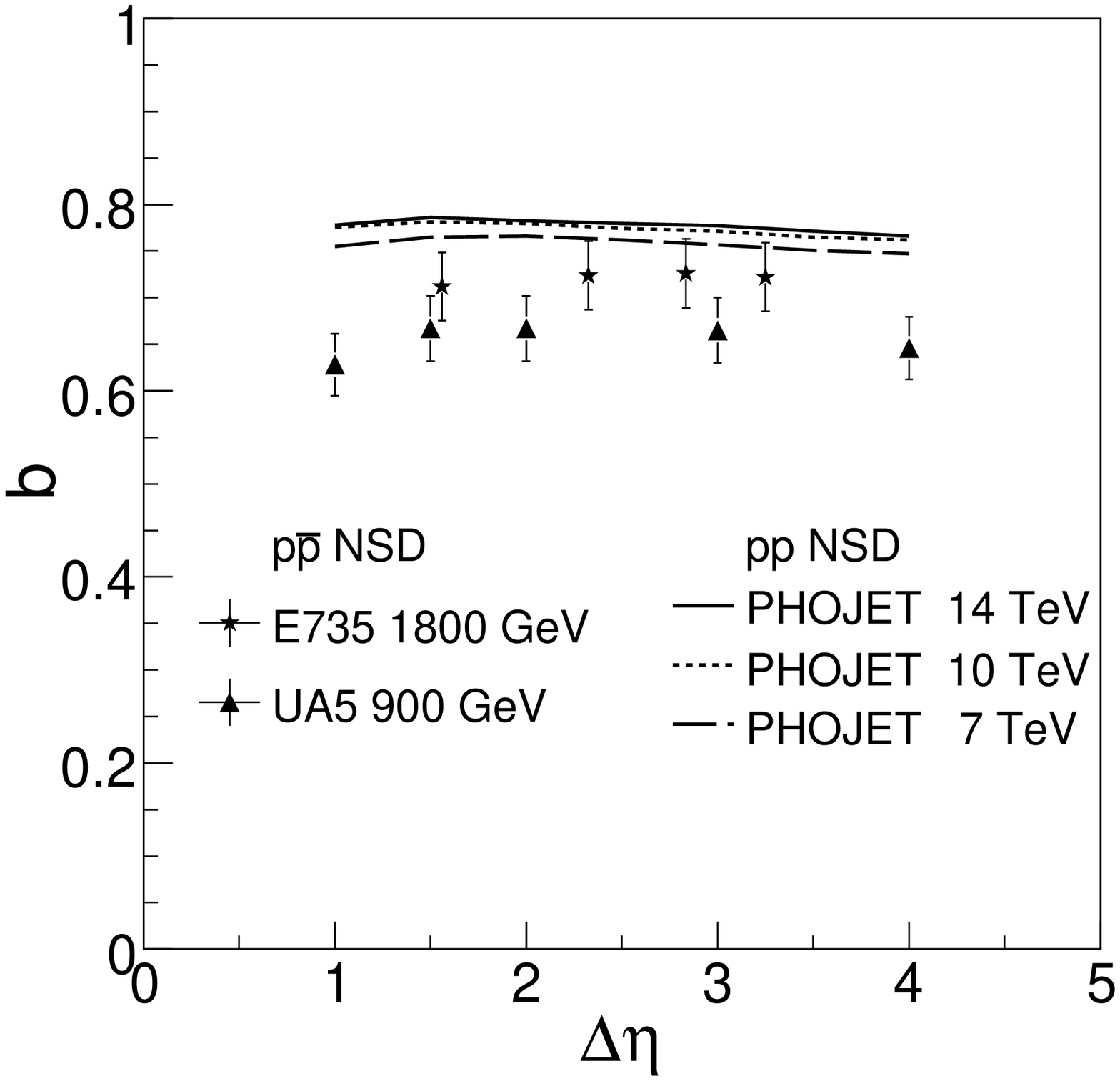}
\caption{Expected forward-backward correlation strength ($b$) estimated
from PYTHIA and PHOJET models as a function of $\Delta\eta$ for $p$+$p$
collisions at LHC energies
of $\sqrt{s}$ = 7, 10 and 14 TeV. Also shown for understanding are the energy
dependence of the correlations from the
existing measurements in $p$+$\bar{p}$ collisions.}
\label{prediction_delta_eta}
\end{center}
\end{figure}

\begin{figure}
\begin{center}
 \includegraphics[scale=0.35]{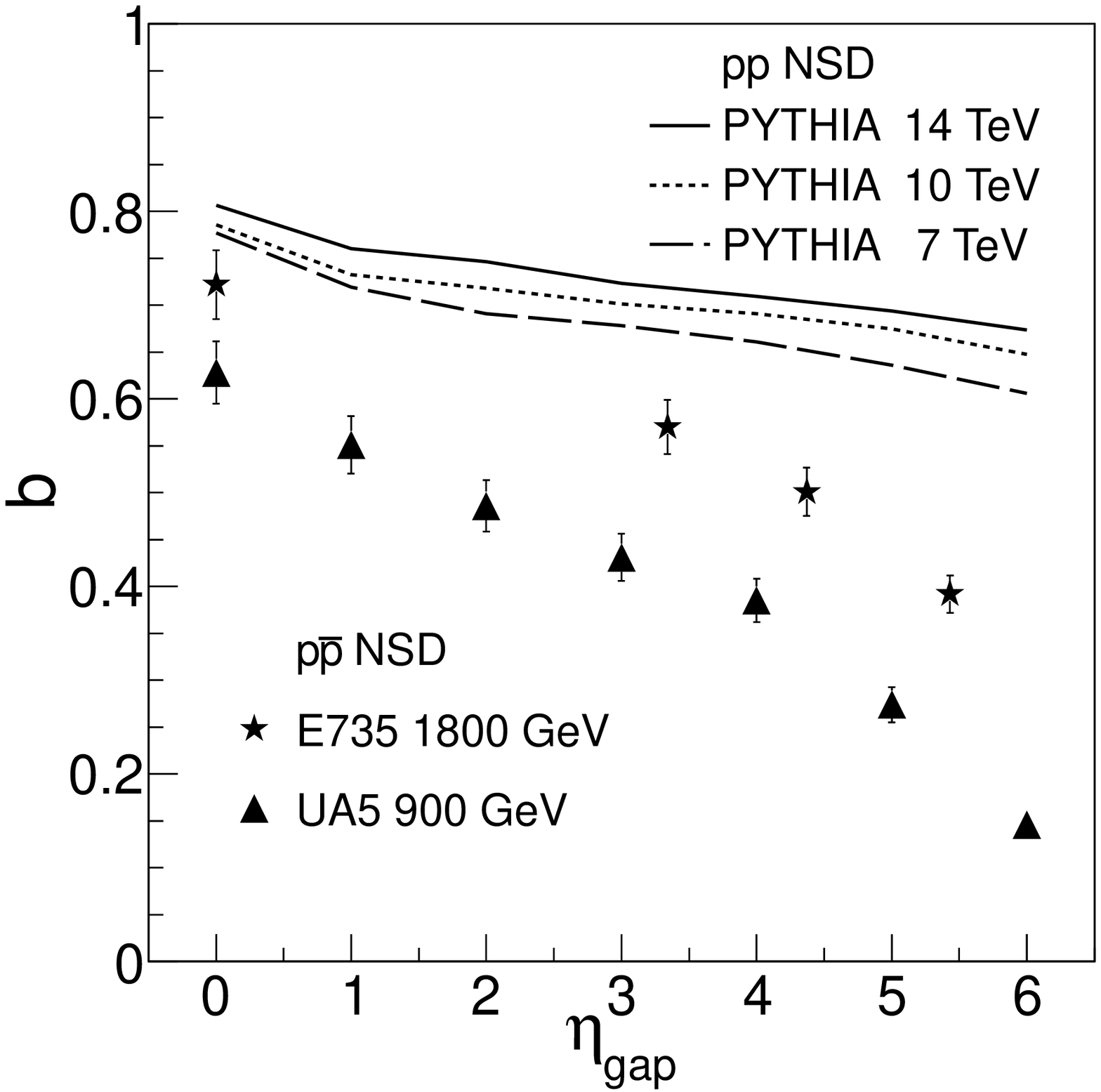}
 \includegraphics[scale=0.35]{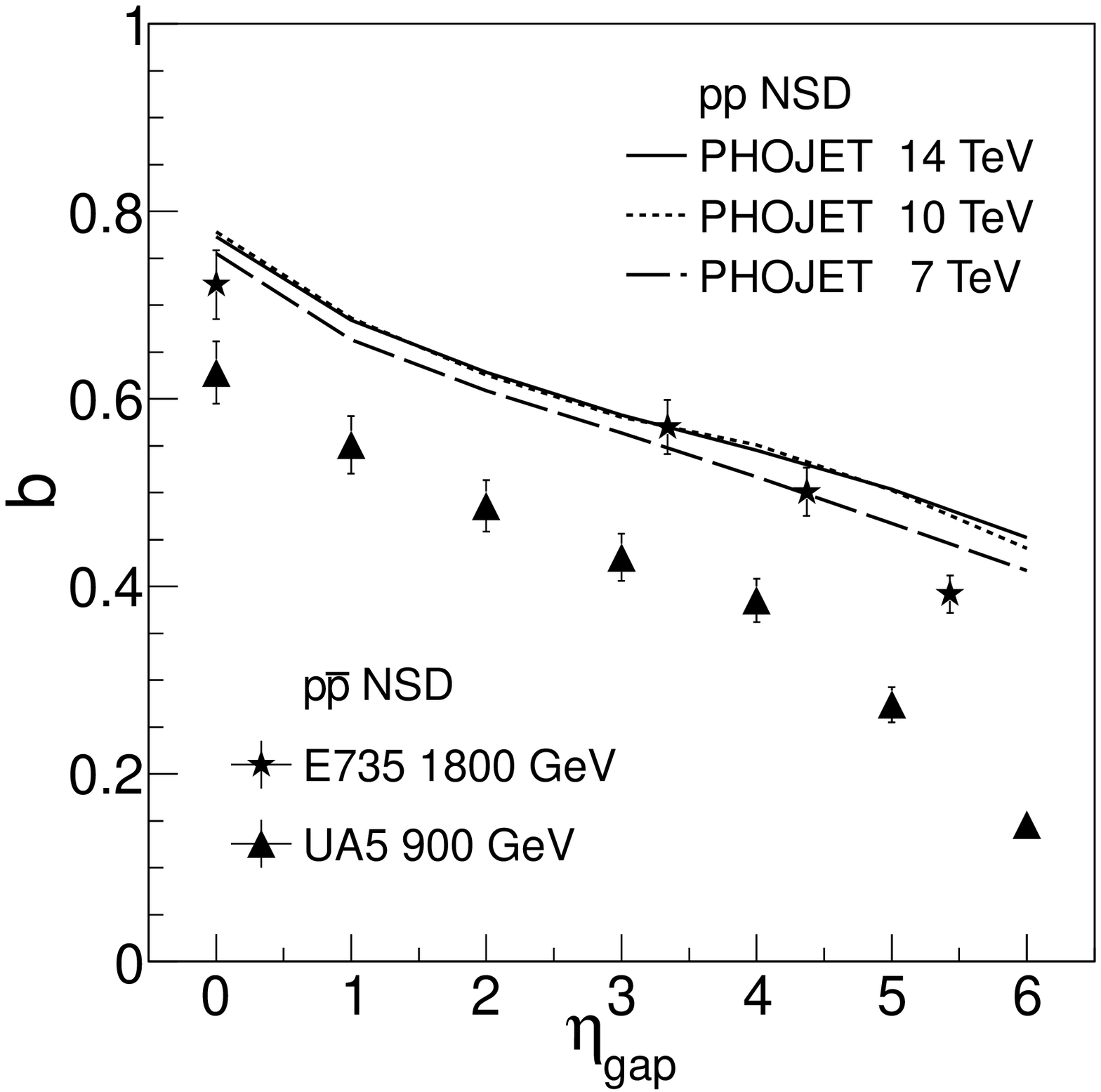}
\caption{Expected forward-backward correlation strength ($b$) estimated
from PYTHIA and PHOJET models as a function of $\eta_{\mathrm {gap}}$ for
$p$+$p$ collisions at LHC energies
of $\sqrt{s}$ = 7, 10 and 14 TeV. Also shown for understanding are the energy
dependence of the correlations for the existing measurements in $p$+$\bar{p}$ collisions.}
\label{prediction_eta_gap}       % Give a unique label                                                                                                                       
\end{center}
\end{figure}

Fig.~\ref{prediction_delta_eta} and ~\ref{prediction_eta_gap} shows
the prediction of forward-backward correlation strength ($b$) from
PYTHIA and PHOJET models as a
function of $\Delta\eta$ and $\eta_{\mathrm {gap}}$ for $p$+$p$
collisions at $\sqrt{s}$ = 7, 10 and 14 TeV. Also shown for understanding
are the energy dependence trend of the existing correlation
measurements from $p$+$\bar{p}$
collisions. In general both models predict a higher but constant value
of $b$ as a function of $\Delta\eta$ at $\sqrt{s}$ = 7, 10 and 14 TeV $p$+$p$
collisions, similar to the trend seen at lower energies $p$+$\bar{p}$
collisions. Although PYTHIA tends to predict a
slightly increasing trend at larger $\Delta\eta$. Both models also predict
a decreasing trend of $b$ as a function of $\eta_{\mathrm {gap}}$ at
 $\sqrt{s}$ = 7, 10 and 14 TeV as was observed for lower energies. The decrease
in  $b$ with respect to $\eta_{\mathrm {gap}}$ from PYTHIA is slower
compared to that from PHOJET at 7, 10 and 14 TeV. Both models clearly suggest
that the correlation strength should saturate at higher energies, with
PHOJET indicating that saturation could occur as early as $\sqrt{s}$ = 1.8 TeV.

\begin{figure}
\begin{center}
\includegraphics[scale=0.6]{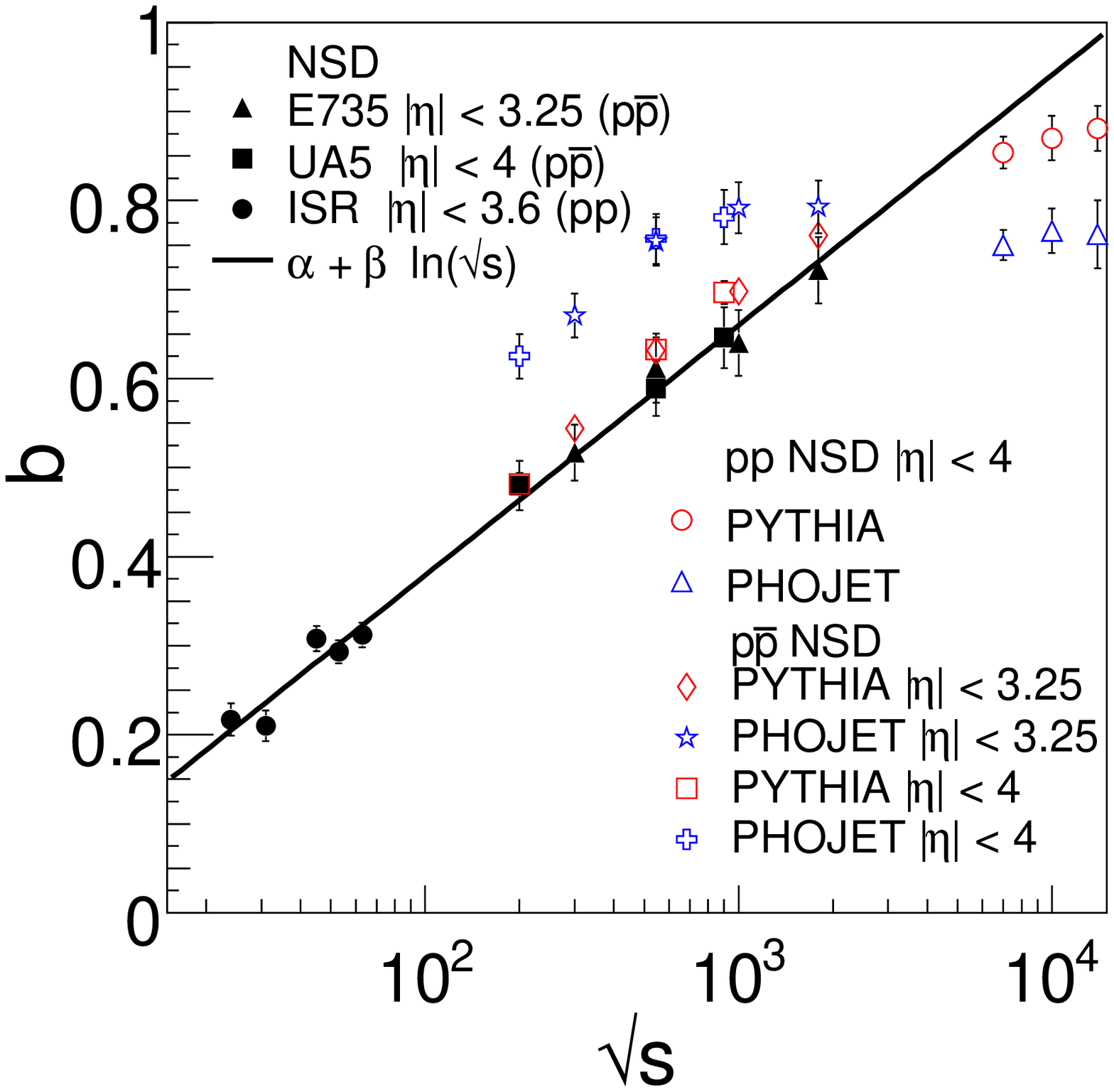}
\caption{Correlation strength, $b$ as a function of ln$\sqrt{s}$ for experimental
data from ISR, UA5 and E735 Collaborations. Also shown are comparisons from PYTHIA
and PHOJET model calculations, including those expected at $\sqrt{s}$
= 7, 10 and 14 TeV.
Solid line is a fit to the experimental data points, extrapolation of which indicates
the correlation strength will reach a maximum value of unity beyond LHC energies.}
\label{energy_dependence}
\end{center}
\end{figure}

The energy dependence of the forward-backward correlation can be
seen in  the Fig.~\ref{energy_dependence}. All available experimental
data from ISR~\cite{isr}, UA5~\cite{ua5} and E735~\cite{e735}
Collaborations shows a linear increase in the correlation value with beam energy.
The best description of the data is obtained as $\alpha$ + $\beta$ln$\sqrt{s}$,
where $\alpha$ = -0.18 $\pm$ 0.02 and $\beta$ = 0.122 $\pm$ 0.005 are the fit parameters.
For the $\sqrt{s}$ $>$ 200 GeV the correlation values from PYTHIA model
are in good agreement with the measurements, but those from PHOJET
model are higher. However interestingly, at the LHC energies the
estimates for $b$ are lower from PYTHIA compared to PHOJET.
If the $\alpha$ + $\beta$ln$\sqrt{s}$  dependence of $b$ on $\sqrt{s}$
holds then the maximum correlation value of 1 will be attained for
$\sqrt{s}$ $\sim$ 16 TeV, beyond LHC energies. Beyond which the value
is expected to saturate. But the model estimates show that the values
of $b$ could saturate or drop earlier around $\sqrt{s}$ = 1.8 TeV.
The PHOJET models predicting a saturation/drop in correlation values
starting at a somewhat smaller $\sqrt{s}$ compared to that predicted
from PYTHIA. New data at LHC energies will provide a clear picture.

Saturation or drop in the energy dependence of the correlation values
could have interesting physical consequences as is discussed in Refs~\cite{clan}.
In this geometric picture, the energy dependence of $b$ is understood based on
the superposition of two components: soft and semi-hard in $p$+$p$($\bar{p}$)
collisions~\cite{clan}. This picture at higher energies is supported by the need of two
negative binomial distributions (with different parameters) to explain the
multiplicity distribution of produced charged hadrons~\cite{ajay}.
Further in such an
approach the
particle production is thought to be from a certain number of independent
initial sources decaying to final products. The produced particles from
each source are expected to all stay in the same hemisphere. However
certain leakage of particles to the other hemisphere is allowed and is
controlled by a leakage parameter in the model. Such a model predicts
a saturation of correlation co-efficient at LHC energies~\cite{clan}.
The final
value of $b$ depends on whether the leakage parameter increases with
$\sqrt{s}$ (higher $b$), constant with $\sqrt{s}$ or decreases with
$\sqrt{s}$ (lower $b$). The prediction from this model agrees well
with the results from PHOJET model and is lower than PYTHIA expectations.

\section{Cluster Production}

One way to interpret the forward-backward correlations is
through the correlations between particles originating
from clusters. The average number of particles originating
from clusters is called the cluster size and can be obtained
by two methods.

\begin{itemize}
\item Method 1: By defining an asymmetry parameter, $Z$ = $N_{f} - N_{b}$,
    then $\langle Z^2 \rangle~=~ r N_{ch}$. Where $r$ is the cluster
    size and $N_{ch}$ is total number of charged particles.
\item Method 2: Directly from the measurement of the correlation coefficient as,
    $b = \frac{\langle N_{ch}\rangle/k + 1 - r}{\langle N_{ch}\rangle/k + 1 + r}$, where  $\langle N_{ch} \rangle$ and $k$ are the parameters of a
   negative binomial fit to the multiplicity distributions~\cite{ajay}.
\end{itemize}

\begin{figure}
\begin{center}
 \includegraphics[scale=0.6]{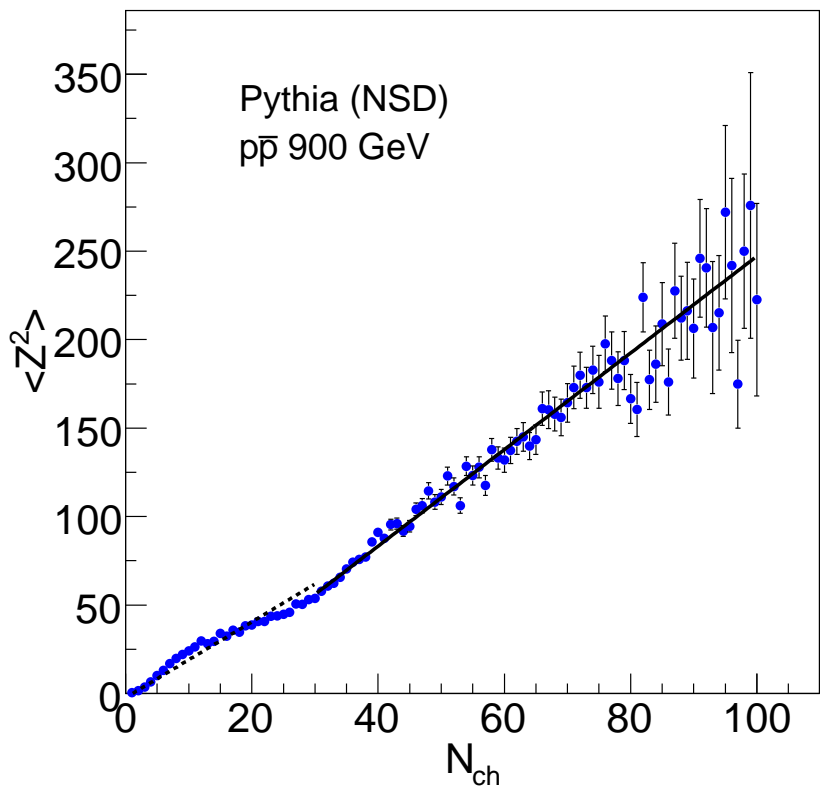}
\caption{Variance of the asymmetry measurement ($\langle Z^2 \rangle$)
as a function of charged particle multiplicity for $p$+$\bar{p}$
collisions at $\sqrt{s}$ = 900 GeV from PYHTIA model. The solid
and dashed lines are a linear fit to extract the cluster sizes in various
$N_{ch}$ ranges.}
\label{Zdist}       % Give a unique label                                                                                                                                    
\end{center}
\end{figure}

Figure~\ref{Zdist} shows the typical relation between $\langle Z^2 \rangle$
and $N_{ch}$ for $p$+$\bar{p}$ collisions at $\sqrt{s}$ = 900 GeV
from PYTHIA. The distribution is fitted by a linear function (dashed and solid lines)
in a fixed interval of $N_{ch}$ to extract the cluster size,$r$.
We followed this procedure to obtain cluster size from both the models
PYTHIA and PHOJET for various $\sqrt{s}$.

\begin{figure}
\begin{center}
 \includegraphics[scale=0.5]{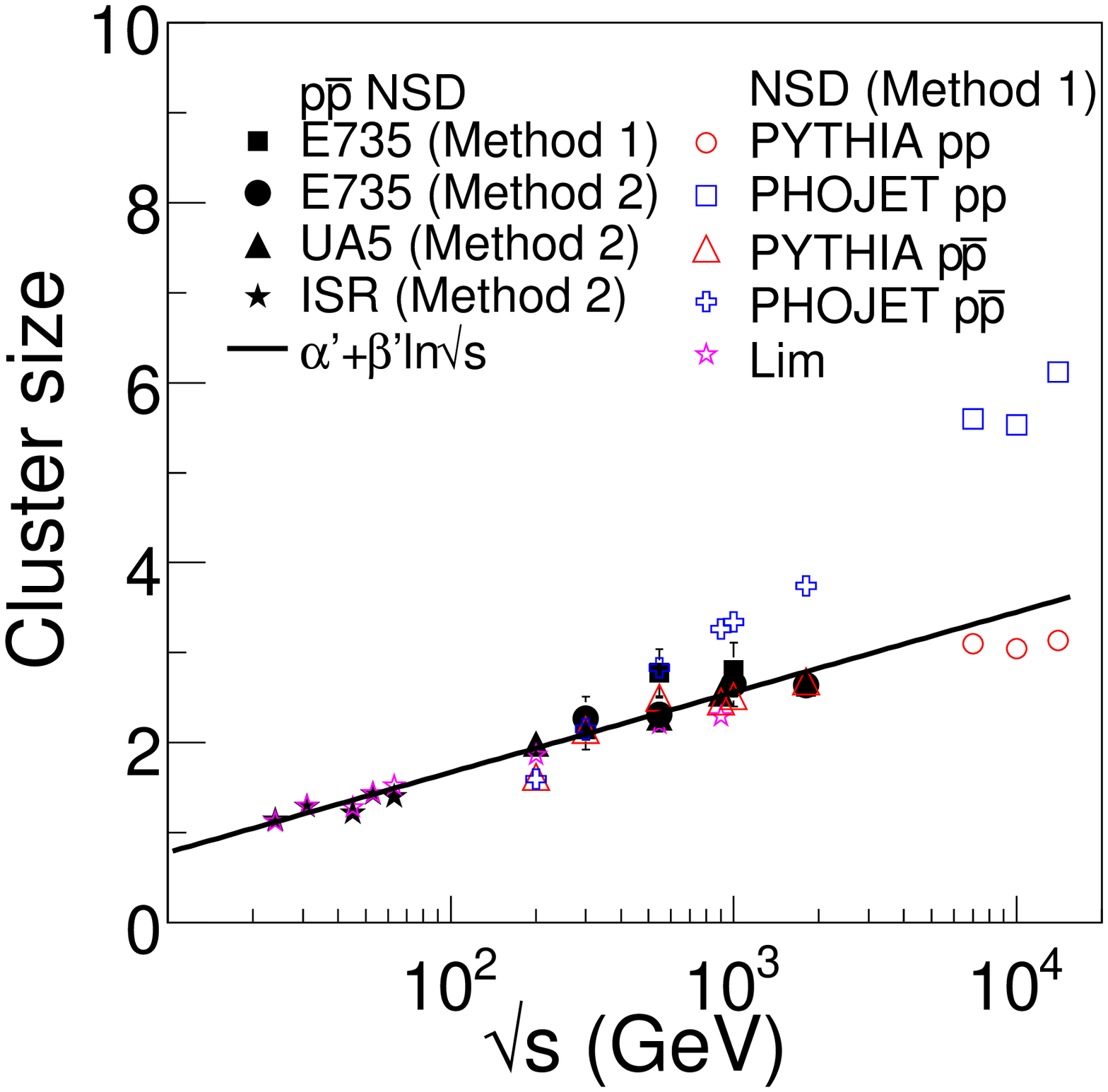}
\caption{Cluster size as a function of $\sqrt{s}$ as measured in
E725, UA5 and ISR experiments.
The method 1 and 2 are discussed in the text.
Also shown are the model comparisons with the data at various beam
energies in p+p collisions and the expectations at top LHC energies
from PYTHIA and PHOJET.}
\label{clustersize}       % Give a unique label                                                                                                                              
\end{center}
\end{figure}

Figure~\ref{clustersize} shows the $\sqrt{s}$ dependence of the cluster
size from E735~\cite{e735}, UA5~\cite{ua5} and ISR~\cite{isr}. The results have been compared to model
calculations from PYTHIA and PHOJET, as well as from a theoretical
calculations by Lim et al~\cite{lim}. The PYTHIA gives similar cluster
size as seen by experiments upto 1.8 TeV and PHOJET give similar
cluster sizes as seen by experiments up to $\sqrt{s}$ = 546 GeV.
Beyond this energy cluster size from PHOJET is higher.
The cluster size dependence on $\sqrt{s}$ for the measured experimental data
can be parameterized as $\alpha^{\prime}$ + $\beta^\prime$ln$\sqrt{s}$. The values
$\alpha^{\prime}$ and $\beta^{\prime}$ for the experimental data, PYTHIA, PHOJET
and from model by Lim et al., are given in the Table.
\begin{table}
\begin{center}
\caption{ Parameters $\alpha^\prime$ and $\beta^\prime$ for the $\sqrt{s}$ dependence of the cluster size.
\label{table}}
\begin{tabular}{ccccc}
\hline
Data/Model & $\alpha^\prime$  & $\beta^\prime$ \\
\hline
Data (UA5, E735, ISR) & -0.11 $\pm$ 0.12  &  0.39 $\pm$ 0.02  \\
Lim                   & 0.12  $\pm$ 0.07  &  0.33 $\pm$ 0.02  \\
PYTHIA                & -0.93 $\pm$ 0.08  &  0.49 $\pm$ 0.01  \\
PHOJET                & -3.7  $\pm$ 0.2   &  1.01 $\pm$ 0.02  \\
\hline
\end{tabular}
\end{center}
\end{table}
The expectations for top LHC energies from the two models are
also shown. PHOJET expects the average number of particles from
a cluster to be around 5-6, while PYTHIA gives a much lower value
of around 3.1. Thereby providing a clear observable, when compared
to LHC data, will distinguish between the underlying
mechanism of particle production in $p$+$p$ collisions.

\section{Summary}

We have reviewed the existing data on forward-backward correlations in
$p$+$\bar{p}(p)$ collisions. Compared these experimental measurements
to model calculations from
PYTHIA and PHOJET. It is found that the correlation strength from PYTHIA
is in agreement with the existing measurements, while those from PHOJET
give higher correlations. However for top LHC energies of $\sqrt{s}$ =
7, 10, 14
TeV, the correlation strength from PHOJET is lower compared to PYTHIA,
suggesting a transition at an intermediate energy accessible at LHC.
The measured correlation strength are found to increase linearly
with $\ln \sqrt{s}$ and extrapolation suggests it will reach
 unity (maximum value) around $\sqrt{s}$ = 20 TeV, beyond
the beam energy reach at LHC. However model calculations suggest the
correlation values tends to saturate starting at $\sqrt{s}$ = 2 TeV. If such
a saturation is observed at LHC it could mean interesting physical consequences
related to clan structures in particle production. We have also reviewed the
existing results on a common interpretation of forward-backward correlations
in terms of cluster production. The cluster sizes are found to increase
with increase in beam energy. Similar to the correlation strength, the
cluster size from PYTHIA compares well with the existing experimental data.
For higher energies ($>$ 546 GeV) PHOJET gives a higher cluster size
compared to PYTHIA. The study of cluster size from forward-backward
correlations can be a very good discriminator for the particle production
models in $p$+$p$ collisions.

\section*{Acknowledgments}

This work is supported by
the DAE-BRNS project sanction No. 2010/21/15-BRNS/2026.

%\section{References}

%\begin{thebibliography}{000} %for 3 digits
%\begin{thebibliography}{00}  %for 2 digits


\begin{thebibliography}{0}    %for 1 digit

%%journal paper
\bibitem{jpap} R. Loren and D. B. Benson, {\it J. Comput. System Sci.} {\bf 27}, 400 (1983).

\bibitem{ALICE} K. Aamodt, {\it et al.}, ALICE Collaboration,
  Eur. Phys. J. C 65 (2010) 111.
\bibitem{ALICE1} K. Aamodt, {\it et al.}, ALICE Collaboration,
  Eur. Phys. J. C 68 (2010) 89.
\bibitem{ALICE2} K. Aamodt, {\it et al.}, ALICE Collaboration,
  Eur. Phys. J. C 68 (2010) 345.

\bibitem{CMS} V. Khachatryan, {\it et al.}, CMS Collaboration,
               JHEP 1002 (2010) 041.

\bibitem{CMS1} V. Khachatryan, {\it et al.}, CMS Collaboration,
               Phys. Rev. Lett. 105 (2010) 022002. 

\bibitem{ATLAS} G. Aad, {\it et al.}, ATLAS Collaboration,
                Phys. Lett. B 688 (2010) 21.

\bibitem{pythia} T.~Sjostrand, {\it et al.}, Computer Physics
  Commun. 135 (2001) 238.

  \bibitem{pythia1}  T.~Sjostrand and M. van Zijl, Phys. Rev. D 36
    (1987) 2019.

\bibitem{pythia2} T.~Sjostrand and P. Skands, Eur. Phys. J. C 39
  (2005) 129.

\bibitem{pythia3} T.~Sjostrand and P. Skands, JHEP 06 (2006) 026.


\bibitem{phojet} R.~Engel, Z. Phys. C 66 (1995) 203.

   \bibitem{phojet1}  R.~Engel, J. Ranft and S. Roesler, Phys. Rev. D 52 (1995) 1459.



\bibitem{dpm} A. Capella, {\it et al.}, Phys. Rep. 236 (1994) 225.

 \bibitem{dpm1}             A. Capella and A. Krzywicki, Phys. Rev. D 18 (1978) 4120.

\bibitem{carruthers} P. Carruthers and C.C. Shih, Phys. Lett. B 165 (1985) 209:

\bibitem{twostage} M.~A. Braun, C. Pajares and V.V. Vechernin, Phys. Lett. B 493 (2000) 54.


\bibitem{dpm2} A. Capella and J. Tran Thanh Van, Z. Phys. C 18 (1983) 85.

\bibitem{minijets} J. Dias de Deus, J. Kwiecinski and M. Pimenta, Phys. Lett. B 202 (1988) 397.

\bibitem{clan} A. Giovannini and R. Ugoccioni, Phys. Rev. D 60 (1999) 074027;
               Phys. Rev. D 59 (1999) 094020.

\bibitem{hwa} R.~C. Hwa and C.~B. Yang, arXiv:0705.3073.

\bibitem{star} B. I. Abelev, {\it et al.}, STAR Collaboration, Phys. Rev. Lett. 103 (2009) 172301.


\bibitem{cgc} Y. V. Kovchegov, E. Levin and L. McLerran,
              Phys. Rev. C 63 (2001) 024903.

\bibitem{cgc1} N. Armesto, L. McLerran and
               C. Pajares, Nucl. Phys. A 781 (2007) 201.

\bibitem{colorsources} P. Brogueira, J. Dias de Deus and C. Pajares,
  Phys. Lett. B 675 (2009) 308.

\bibitem{e735} T. Alexopoulos, {\it et al.}, E735 Collaboration, Phys. Lett. B 353 (1995) 155.

\bibitem{ua5}R. E. Ansorge, {\it et al.}, UA5 Collaboration,
 Z. phys. C 37 (1988) 191.


\bibitem{brijesh} B.K. Srivastava, STAR Collaboration,
  Int. J. Mod. Phys. E 16 (2008) 3371.


\bibitem{isr}S. Uhlig, {\it et al.}, Nucl. Phys. B 132 (1978) 15.

\bibitem{ajay} A. K. Dash and B. Mohanty, J. Phys. G 37 (2010) 025102.


\bibitem{lim}S.~L.~ Lim {\it et al.}, Z. Phys. C 43 (1989) 621.

\end{thebibliography}
\end{document}